# The effect of a country's name in the title of a publication on its visibility and citability[1]

*Giovanni Abramo*[a,c], *Ciriaco Andrea D'Angelo*[b,a], *Flavia Di Costa*[a]

[a] Laboratory for Studies of Research and Technology Transfer
Institute for System Analysis and Computer Science (IASI-CNR)
National Research Council of Italy

[b] Department of Engineering and Management
University of Rome "Tor Vergata"

**Abstract**

The objective of this research is to determine if the reference to a country in the title, keywords or abstract of a publication can influence its visibility (measured by the impact factor of the publishing journal) and citability (measured by the citations received). The study is based on Italian scientific production indexed in the Web of Science over the period 2004-2011. The analysis is conducted by comparing the values of four impact indicators for two subsets: i) the indexed publications with a country's name in the title, keywords or abstract; ii) the remainder of the population, with no country' name. The results obtained both at the general level and by subject category show that publications with a country name systematically receive lower impact values, with the exception of a limited number of subject categories, Also, the incidence of highly-cited articles is lower for the first subset.

# 1. Introduction

Many empirical studies have been conducted to identify factors unrelated to the content of an article which could influence its visibility and citability. Bornmann and Daniel (2008), in their review of studies on citing behavior, identify two general schools of thought among the scholars in the field. One group holds that the numbers of citations to publications are influenced primarily by the personal prestige of the authors and of their institutions of affiliation (e.g. departmental prestige, research grants, academic rank, awards, Nobel prizes, other honors, peer judgments). The other group instead considers that the number of citations is influenced by a multitude of factors (e.g. author/reader-dependent factors, time-dependent factors; field- and journal-dependent factors; availability of publications; technical problems).

Van Wesel et al., (2014) focus their attention on the length of titles and abstracts, numbers of pages, authors and cited references, readability, etc. Various studies have concentrated on the importance of the article title, since as Haggan (2004) reasons, "the title plays an important role as the first point of contact between writer and potential reader and may decide whether or not the paper is read." These studies can be divided in three classes of approach or interest: i) works on the formal syntactic structures of titles (Wang & Bai, 2007; Cheng et al., 2012; Salager-Meyer & Ariza, 2013; Wang et al., 2014; Bavdekar, 2016); ii) studies on the relationship between the structure of the title and visibility (Haggan, 2004; Demner-Fushman et al., 2005; Demner-Fushman et al., 2005; Morrison & Batty, 2009); iii) works on the relation between the structure of the title and citation rates (Habibzadeh & Yadollahie, 2010; Jacques & Sebire, 2010; Jamali & Nikzad, 2011; Paiva et al., 2012; Subotic & Mukherjee, 2014; Rostami et al., 2014; Falahati et al., 2015).

Bavdekar (2016) limits the typology of titles to three broad classes: declarative, informative and interrogative. *Declarative titles* present the main results or conclusions contained in the paper; *Descriptive titles* are limited to describing the theme of the article; *Interrogative titles* generally reformulate the research question. From a structural perspective, paper titles can be classified as nominal, compound or full-sentence titles. *Nominal titles* present the general premise of the study; *Compound titles* (or "hanging titles") have a subtitle, used to provide further information relevant to the study (e.g. context, research design). Finally there are *full sentence titles,* which are the longest, containing greater information on the results obtained by the study (e.g. highlighting a central result). Cheng et al. (2012) have examined the syntactic structures and function of research article titles in the field of applied linguistics, while Hartley (2012)[2] has identified a full 13 types of studies based on the contents and purpose of the works.

Most of the studies focus on the relations between the structure of the title and the impact of the publication in citational terms. Falahati et al. (2015) have conducted a morphological analysis of titles, for study of the link between citability and title length/number of punctuation marks. The results of the analysis, conducted on a sample of 650 articles published in the journal *Scientometrics* over the years 2009-2011 show that: i) title length and citations to articles are not correlated; ii) the variable of number of punctuation marks does not serve as a reliable predictor of an article's citation results. Habibzadeh and Yadollahie (2010) have also studied the correlation between the length of an article title and the number of citations, for the particular area of the medical sciences. Their results demonstrate that longer titles seem associated with higher citation rates, above all for articles published in journals with high impact factors (IF). Using a sample including all the articles published in six PLOS journals, Jamali and Nikzad (2011) have investigated the influence of the type of article title on the number of citations and downloads that an article receives. They observe that: i) "question" articles tend to be downloaded more often, but cited less compared to others; ii) articles with longer titles are downloaded less than those with shorter titles; iii) titles with colons tend to be longer, and therefore receive less downloads and citations. Rostami et al. (2014) have studied the association between some features of titles relative to numbers of citations, examining the articles of the 2007 volume of *Addictive Behavior*: the results indicate that the type of title, as well as the number of keywords different from the words in the title, can contribute to predicting the number of citations for the publications.

---

[2] http://blogs.lse.ac.uk/impactofsocialsciences/2012/05/24/titles-are-hardest-more-effective/, last access July 11, 2016.

To the best of our knowledge, only three studies have included any consideration of the effect of the presence of a country name in the publication title on the citation rate for the work: those of Jacques and Sebire, (2010), Paiva et al. (2012), and Nair and Gibbert (2016).

The first of these studies examines the title characteristics of the 25 most-cited articles and the 25 least-cited articles published in general and specialist medical journals in the year 2005. The results show that the number of citations is positively correlated with the length of the title, the presence of a colon in the title and the presence of an acronym. Furthermore, the reference to a specific country in the title is among those factors that predict poor citation. In fact the authors state: "The most striking illustration of the effects of title words, however, appears to be the dramatic adverse effect of having a specific country mentioned in the title, this being present in more than one-third of poorly-cited Lancet articles but none of the well cited articles, despite many of the well-cited group representing specific studies performed in single countries. It is likely that when searching for evidence, many researchers may discount information, which is perceived to only relate to another specific country".

In the second study, Paiva et al. (2012) analyze the titles of 423 articles published in open access journals in 2008 (PLOS and Biomed Central), verifying the possible impact of titles as predictors of the number of article views and citations. The results obtained show that the works with titles containing a reference to a specific geographical region receive a significantly lower number of citations.

In the third study, Nair & Gibbert (2016) examine whether title attributes providing specific contextual information influence citation count. They analyze a sample of 553 titles out of 2597 articles published over a decade in the five top tier management journals. The authors find that title's length, context and linguistic attributes have no relationship with citation count. Within the term context, in addition to the country name, the authors include also two more management specific context attributes, company name and industry name.

The common traits of the above three studies is that they investigate specific research fields, and base their conclusions on a quite limited number of observations. The objective of the present study is to overcome these limits. We examine whether a country's name (in this case "Italy", or a related adjective) in the title, abstract or keywords of a publication could be a penalizing factor in terms of the visibility, measured by the IF of the publishing journal, and impact, as measured by the number of citations received. The analysis will be conducted at the general level and by subject categories, to detect any potential sectoral differences. The study will also examine at the deeper level of the subcategory of highly-cited articles. The added value represented by the study lies in: i) the scope and character of the dataset examined, consisting of the entire scientific production of a given nation (456,710 publications), over a sufficiently extensive time interval (2004-2011); ii) the breadth of scientific disciplines analyzed, consisting of all the WoS subject categories in the sciences (171), social sciences (50), arts and humanities (27); and multidisciplinary sciences (3) and iii) the number of bibliometric indicators used (four), regarding both the impact factor of the publishing journal and the citations received by the works (measured on 31/05/2014).

The next section presents the methodology used in the study and the description of the dataset. Section 3 reports the results of the analyses, conducted both at the general level and by subject category, with the further specific focus on the rate of concentration of Highly Cited Articles (HCAs). The final section offers the conclusions, with some considerations on the limits of the study and potential future developments.

## 2. Methodology and dataset

The study utilizes the following indicators of impact and visibility of the publications:

*Article Impact Index, AII*. The ratio of the numbers of citations received by the publication to the average of the citations for all national publications cited[3] of the same year and subject category[4].

*Article Impact Rank, AIR*. The publication percentile rank by citations, obtained by comparison with all national publications of the same year and subject category[3].

*Journal Impact Index, JII*. The ratio of the IF of the journal to the average IF of the journals of the same subject category.

*Journal Impact Rank, JIR*. The publication percentile rank by IF, obtained by comparison with all journals of the same subject category.

The field of observation consists of all publications indexed in the Italian National Country Report extracted from the core collection of the Web of Science, and made available to the authors by Thomson Reuters under a license agreement. The analysis concerns all publications indexed in the period 2004-2011 (identifying 456,710 records in all). Next, we divided the total publications of the dataset into two subsets:

- The first, named "Country", numbering 40,024 records, includes all publications having at least one of the terms *Italy*, *Italian*, *Italic* in any of the bibliographic record fields Abstract, Keywords or Title.
- The second, named "No country", includes the remaining 416,686 records lacking the country's name.

Table 1 presents the distribution of the publications per document type in the two subsets. A first observation is that for the Country subset there is a greater incidence of "article" documents. Table 2 presents the distribution of the three terms referring to country (*Italy, Italian, Italic*), per document type and bibliographic field. We observe that the fields where the terms are most present are, in order: Abstract, Title, and finally

---
[3] Concerning the choice of the scaling factor, see Abramo et al. (2012).
[4] For publications in multi-category journals, the value of the indicator is calculated as the average of the values for the individual subject categories.

Keywords. The total of 60,404 observations includes repeat counts, since any of the three terms can appear simultaneously in one or more fields of a given bibliographic record. The data reported in the last row of Table 2 exclude these multiple counts.

*Table 1: Distribution of document types in the two subsets*

| Document type | "Country" | "No Country" |
|---|---|---|
| Article | 33,599 (83.9%) | 321,538 (77.2%) |
| Proceedings paper | 4,583 (11.5%) | 52,216 (12.5%) |
| Review | 1,014 (2.5%) | 23,269 (5.6%) |
| Letter | 540 (1.3%) | 16,089 (3.9%) |
| Others | 288 (0.7%) | 3,574 (0.9%) |
| Total | 40,024 | 416,686 |

*Table 2: Distribution of the reference words in the fields per document type*

| | | Document type | | | | | |
|---|---|---|---|---|---|---|---|
| | Country word | Article | Proceedings Paper | Review | Letter | Other | Total |
| Title | Italy | 9,514 | 930 | 267 | 231 | 160 | 11,102 (63.1%) |
| | Italian | 5,390 | 539 | 140 | 210 | 134 | 6,413 (36.5%) |
| | Italic | 70 | 3 | | | 2 | 75 (0.4%) |
| | *Total* | *14,974* | *1,472* | *407* | *441* | *296* | *17,590 (100%)* |
| Abstract | Italy | 19,470 | 2,604 | 507 | 13 | 1 | 22,595 (58.3%) |
| | Italian | 13,547 | 1,887 | 441 | 90 | 1 | 15,966 (41.2%) |
| | Italic | 179 | 15 | 2 | | | 196 (0.5%) |
| | *Total* | *33,196* | *4,506* | *950* | *103* | *2* | *38,757 (100.0%)* |
| Keywords | Italy | 2,655 | 110 | 100 | 7 | 1 | 2,873 (70.8%) |
| | Italian | 999 | 46 | 65 | 8 | 1 | 1,119 (27.6%) |
| | Italic | 63 | 2 | | | | 65 (1.6%) |
| | *Total* | *3,717* | *158* | *165* | *15* | *2* | *4,057 (100.0%)* |
| | *Total* | *51,887* | *6,136* | *1,522* | *559* | *300* | *60,404* |
| *Total without duplicates* | | *33,599* | *4,583* | *1,014* | *540* | *288* | *40,024* |

## 3. Analysis and results

### 3.1 General analysis

Table 3 presents the descriptive statistics for the four impact indicators calculated on the two subsets of publications. We observe that with all of the indicators, the values for the average, standard deviation and maximum of the No country subset are greater than those for the Country subset. Table 4 replicates the previous one, except that in this case the descriptive statistics for the two subsets are observed in function of the document type. The average values of the four indicators are greater for the No country than Country subset for all types of documents, except for the average AIR and AII of the "Letter" type.

Next we focus the comparison between the publications with the country's name in the title only and those without. We exclude then from the "Country" subset the publications with the reference to the country in the abstract and in the keywords, which are now included in the "No country" subset. We call these subsets "Country (t)" and "No country (t)". The results reported in Table 5 and Table 6 show that the impact differences are even higher than in the previous analysis.

*Table 3: Descriptive statistics of impact indicators for the "Country" and "No country" subsets*

|  | Country | | | No country | | |
|---|---|---|---|---|---|---|
| Obs | 40,024 | | | 416,686 | | |
| Indicator | Mean | Std Dev. | Max | Mean | Std Dev. | Max |
| JIR | 47.0 | 33.4 | 100 | 54.1 | 34.2 | 100 |
| JII | 0.924 | 0.918 | 21.1 | 1.151 | 1.344 | 22.6 |
| AIR | 47.4 | 30.7 | 100 | 49.3 | 32.0 | 100 |
| AII | 0.709 | 1.075 | 35.5 | 0.846 | 1.751 | 204.2 |

*Legend: JIR = publication percentile rank by IF; JII = field-normalized IF; AIR = publication percentile rank by citations; AII = field-normalized citations*

*Table 4: Descriptive statistics of impact indicators per document type for the "Country" and "No country" subsets*

| Subset | Document Type | Obs | Mean | | | | Std. Dev. | | | | Max | | | |
|---|---|---|---|---|---|---|---|---|---|---|---|---|---|---|
| | | | *JIR* | *JII* | *AIR* | *AII* | *JIR* | *JII* | *AIR* | *AII* | *JIR* | *JII* | *AIR* | *AII* |
| Country | Article | 33,599 | 53.1 | 1.037 | 52.4 | 0.795 | 30.4 | 0.876 | 28.3 | 1.072 | 100 | 21.070 | 100 | 34.9 |
| | Review | 1,014 | 53.0 | 1.061 | 57.0 | 1.160 | 32.4 | 0.945 | 29.9 | 2.183 | 100 | 9.849 | 100 | 35.5 |
| | Proceedings Paper | 4,583 | n.a. | n.a. | 14.0 | 0.077 | n.a. | n.a. | 24.0 | 0.218 | n.a. | n.a. | 97.9 | 3.2 |
| | Letter | 540 | 71.7 | 1.759 | 31.0 | 0.289 | 24.6 | 1.976 | 25.6 | 0.427 | 100 | 17.372 | 95.0 | 2.8 |
| | Other | 288 | 16.5 | 0.324 | 1.0 | 0.004 | 29.4 | 0.692 | 9.3 | 0.049 | 100 | 5.432 | 98.6 | 0.8 |
| No country | Article | 321,538 | 61.6 | 1.265 | 55.2 | 0.935 | 29.0 | 1.155 | 28.7 | 1.605 | 100 | 21.070 | 100 | 153.1 |
| | Review | 23,269 | 64.1 | 1.498 | 66.2 | 1.882 | 30.2 | 1.641 | 27.9 | 3.994 | 100 | 22.581 | 100 | 204.2 |
| | Proceedings Paper | 52,216 | n.a. | n.a. | 16.8 | 0.105 | n.a. | n.a. | 27.9 | 0.327 | n.a. | n.a. | 99.9 | 18.6 |
| | Letter | 16,089 | 72.2 | 2.264 | 21.1 | 0.165 | 26.3 | 3.143 | 22.6 | 0.395 | 100 | 21.070 | 99.8 | 14.1 |
| | Other | 3,574 | 18.5 | 0.407 | 1.8 | 0.038 | 29.3 | 1.118 | 11.7 | 1.047 | 100 | 15.688 | 100 | 58.9 |

*Legend: JIR = publication percentile rank by IF; JII = field-normalized IF; AIR = publication percentile rank by citations; AII = field-normalized citations*

*Table 5: Descriptive statistics of impact indicators for the "Country (t)" and "No country (t)" subsets*

|           | Country's name in the title | | | No country's name in the title | | |
|-----------|------|---------|-----|------|---------|-----|
| Indicator | Mean | Std dev. | Max | Mean | Std dev. | Max |
| JIR | 46.3 | 32.8 | 100 | 53.7 | 34.2 | 100 |
| JII | 0.881 | 0.796 | 17.4 | 1.138 | 1.324 | 22.6 |
| AIR | 46.6 | 29.7 | 100 | 49.2 | 31.9 | 100 |
| AII | 0.656 | 0.829 | 17.3 | 0.839 | 1.719 | 204.2 |

*Legend: JIR = publication percentile rank by IF; JII = field-normalized IF; AIR = publication percentile rank by citations; AII = field-normalized citations*

*Table 6: Descriptive statistics of impact indicators per document type for the "Country (t)" and "No country (t)" subsets*

| Subset | Document Type | Mean | | | | Std dev. | | | | Max | | | |
|--------|---------------|------|-----|-----|-----|------|------|------|------|------|------|------|------|
|        |               | JIR | JII | AIR | AII | JIR | JII | AIR | AII | JIR | JII | AIR | AII |
| Country's name in title | Article | 50.5 | 0.945 | 50.5 | 0.717 | 30.6 | 0.665 | 27.7 | 0.838 | 100 | 9.3 | 100 | 17.3 |
| | Review | 56.5 | 1.105 | 57.7 | 1.053 | 32.0 | 0.786 | 30.4 | 1.172 | 100 | 4.3 | 99.6 | 8.7 |
| | Proceedings Paper | n.a. | n.a. | 15.6 | 0.110 | n.a. | n.a. | 24.1 | 0.291 | n.a. | n.a. | 97.9 | 3.2 |
| | Letter | 70.0 | 1.913 | 28.7 | 0.301 | 26.8 | 2.567 | 26.3 | 0.497 | 100 | 17.4 | 95.0 | 2.8 |
| | Other | 17.1 | 0.327 | 1.2 | 0.007 | 29.2 | 0.668 | 10.4 | 0.066 | 100 | 5.4 | 98.6 | 0.8 |
| No Country's name in title | Article | 61.1 | 1.252 | 55.1 | 0.927 | 29.1 | 1.142 | 28.7 | 1.578 | 100 | 21.1 | 100 | 153.1 |
| | Review | 63.7 | 1.483 | 65.9 | 1.860 | 30.4 | 1.627 | 28.0 | 3.957 | 100 | 22.6 | 100 | 204.2 |
| | Proceedings Paper | n.a. | n.a. | 16.6 | 0.102 | 4.0 | 0.082 | 27.7 | 0.320 | 97.8 | 3.1 | 99.9 | 18.6 |
| | Letter | 72.2 | 2.252 | 21.3 | 0.167 | 26.2 | 3.121 | 22.7 | 0.395 | 100 | 21.1 | 99.8 | 14.1 |
| | Other | 18.4 | 0.404 | 1.8 | 0.037 | 29.4 | 1.107 | 11.6 | 1.029 | 100 | 15.7 | 100 | 58.9 |

*Legend: JIR = publication percentile rank by IF; JII = field-normalized IF; AIR = publication percentile rank by citations; AII = field-normalized citations*

### 3.2 Analyses by subject category

To detect any differences in behavior between subject categories, we analyze the publications of the Country subset stratified for this factor. Because of space limitations, we report the analysis of the 20 subject categories that show the highest shares of publications with country terms in the metadata. The subject category that shows the highest percentage is KY_Geology (51.7%), followed by TE_Paleontology (51.2%) and KV_Geography, physical (50.1%). After having presented the descriptive statistics for each impact indicator in each subject category, we will discuss the significance of the differences that are encountered.

#### 3.2.1 Descriptive statistics

Tables 7 and 8 present the descriptive statistics for the indicators JIR, JII, AIR and AII for each of the 20 subject categories considered. Given that many proceedings papers are not assigned an impact factor, we have excluded these from consideration in calculating the descriptive statistics for JIR and JII.
- Concerning JIR and JII, the average value for the Country subset is greater than for the No country subset in only three subject categories (IM_Engineering, civil; LE_Geosciences, multidisciplinary; ZC_Veterinary sciences).

- Concerning AIR, the average value of the Country subset is greater than the average of the No country subset in six subject categories (DM_Oncology; LE_Geosciences, multidisciplinary; MU_Horticulture; NE_Public, environmental & occupational health; NN_Infectious diseases; QU_Microbiology; ZC_Veterinary sciences).
- Concerning AII, the average value of the Country subset is greater than for the No country subset in only three subject categories (LE_Geosciences, multidisciplinary; QU_Microbiology; ZC_Veterinary sciences).

*Table 7: Descriptive statistics for the JIR and AIR indicators, for 20 subject categories*

| Subject category* | Indicator** | Country | | | | No Country | | | |
|---|---|---|---|---|---|---|---|---|---|
| | | Obs | Mean | Std Dev. | Max | Obs | Mean | Std Dev. | Max |
| DE | JIR | 928 | 47.3 | 23.6 | 96.8 | 3,868 | 64.1 | 27.5 | 100 |
| | AIR | 1,253 | 40.8 | 28.2 | 100 | 4,353 | 53.6 | 29.6 | 100 |
| DM | JIR | 1,299 | 58.6 | 29.9 | 100 | 15,608 | 63.6 | 29.2 | 100 |
| | AIR | 1,304 | 52.7 | 27.0 | 99.6 | 15,675 | 52.2 | 29.0 | 100 |
| GC | JIR | 1,688 | 48.5 | 34.6 | 100 | 3,311 | 54.4 | 35.1 | 100 |
| | AIR | 1,801 | 51.1 | 28.7 | 100 | 3,472 | 54.0 | 30.5 | 100 |
| GU | JIR | 846 | 57.1 | 28.9 | 100 | 2,395 | 60.7 | 28.3 | 100 |
| | AIR | 937 | 49.8 | 29.4 | 99.2 | 2,554 | 52.7 | 29.9 | 100 |
| IM | JIR | 543 | 62.4 | 29.6 | 100 | 2,339 | 61.5 | 26.8 | 100 |
| | AIR | 906 | 40.7 | 35.1 | 99.8 | 3,371 | 46.9 | 34.2 | 100 |
| JA | JIR | 2,402 | 58.3 | 30.1 | 100 | 6,136 | 61.3 | 31.1 | 100 |
| | AIR | 2,765 | 48.7 | 29.9 | 100 | 6,826 | 51.7 | 30.3 | 100 |
| KV | JIR | 769 | 61.2 | 28.0 | 99.6 | 753 | 65.2 | 26.4 | 99,6 |
| | AIR | 785 | 54.0 | 29.3 | 99.9 | 783 | 56.6 | 28.9 | 100 |
| KY | JIR | 603 | 43.7 | 34.1 | 100 | 566 | 52.8 | 34.8 | 100 |
| | AIR | 751 | 47.5 | 30.4 | 100 | 703 | 50.8 | 31.4 | 100 |
| LE | JIR | 2,804 | 61.1 | 27.0 | 100 | 4,108 | 58.9 | 29.0 | 100 |
| | AIR | 3,067 | 54.2 | 28.1 | 99.9 | 4,867 | 50.4 | 30.6 | 100 |
| MU | JIR | 236 | 57.9 | 20.7 | 92.0 | 826 | 62.9 | 22.8 | 93,0 |
| | AIR | 982 | 31.1 | 32.8 | 99.2 | 1,951 | 40.7 | 34.7 | 100 |
| NE | JIR | 1,456 | 50.4 | 32.1 | 100 | 2,745 | 52.8 | 33.2 | 100 |
| | AIR | 1,492 | 51.8 | 28.0 | 100 | 2,923 | 48.5 | 31.3 | 100 |
| NN | JIR | 859 | 60.3 | 26.5 | 100 | 3,110 | 67.9 | 26.1 | 100 |
| | AIR | 880 | 52.4 | 25.8 | 99.8 | 3,156 | 50.0 | 29.8 | 100 |
| PI | JIR | 721 | 55.6 | 25.5 | 98.9 | 2,202 | 61.2 | 24.5 | 100 |
| | AIR | 738 | 49.8 | 27.1 | 98.7 | 2,222 | 52.5 | 27.8 | 100 |
| QQ | JIR | 786 | 53.1 | 28.1 | 100 | 2,525 | 56.8 | 32.9 | 100 |
| | AIR | 872 | 49.2 | 28.0 | 98.2 | 2,748 | 53.8 | 29.7 | 100 |
| QU | JIR | 880 | 61.6 | 26.0 | 98.9 | 4,477 | 63.5 | 26.3 | 100 |
| | AIR | 888 | 55.5 | 27.7 | 99.9 | 4,505 | 52.7 | 28.3 | 100 |
| RE | JIR | 427 | 52.3 | 31.9 | 96.8 | 935 | 62.5 | 26.0 | 99,1 |
| | AIR | 441 | 47.5 | 26.6 | 99.9 | 951 | 52.0 | 27.6 | 100 |
| TE | JIR | 502 | 47.9 | 30.9 | 98.4 | 490 | 58.7 | 29.4 | 98,4 |
| | AIR | 525 | 49.2 | 26.9 | 98.3 | 500 | 54.5 | 26.8 | 100 |
| ZC | JIR | 994 | 48.0 | 32.0 | 100 | 4,127 | 45.8 | 30.2 | 100 |
| | AIR | 1,022 | 51.8 | 31.0 | 99.6 | 4,164 | 46.3 | 31.4 | 100 |
| ZM | JIR | 661 | 36.6 | 23.4 | 98.6 | 2,141 | 46.8 | 25.7 | 100 |
| | AIR | 666 | 44.0 | 28.2 | 96.8 | 2,151 | 49.0 | 27.7 | 100 |
| ZR | JIR | 1,049 | 58.0 | 27.3 | 96.5 | 1,998 | 63.0 | 26.6 | 96,9 |
| | AIR | 1,217 | 48.9 | 28.8 | 99.7 | 2,298 | 50.7 | 30.4 | 100 |

* DE=Plant Sciences; DM=Oncology; GC=Geochemistry & Geophysics; GU=Ecology; IM=Engineering, Civil; JA=Environmental Sciences; KV=Geography, Physical; KY=Geology; LE=Geosciences, Multidisciplinary; MU=Horticulture; NE=Public, Environmental & Occupational Health; NN=Infectious Diseases; PI=Marine & Freshwater Biology; QQ=Meteorology & Atmospheric

Sciences; QU=Microbiology; RE=Mineralogy; TE=Paleontology; ZC=Veterinary Sciences; ZM=Zoology; ZR=Water Resources
** JIR = publication percentile rank by IF; AIR = publication percentile rank by citations

*Table 8: Descriptive statistics for the JII and AII indicators, for 20 subject categories*

| Subject category* | Indicators** | Country | | | | No country | | | |
|---|---|---|---|---|---|---|---|---|---|
| | | Obs | Mean | Std Dev. | Max | Obs | Mean | Std Dev. | Max |
| DE | JII | 928 | 0.75 | 0.52 | 3.50 | 3,868 | 1.36 | 1.11 | 15.69 |
| | AII | 1,253 | 0.51 | 0.83 | 15.74 | 4,353 | 0.96 | 1.31 | 25.18 |
| DM | JII | 1,299 | 1.14 | 1.27 | 21.07 | 15,608 | 1.32 | 1.51 | 22.58 |
| | AII | 1,304 | 0.81 | 1.23 | 13.67 | 15,675 | 1.00 | 2.63 | 104.45 |
| GC | JII | 1,688 | 0.95 | 0.94 | 14.47 | 3,311 | 1.17 | 1.38 | 17.37 |
| | AII | 1,801 | 0.83 | 1.05 | 11.55 | 3,472 | 0.99 | 1.38 | 29.01 |
| GU | JII | 846 | 1.10 | 0.85 | 15.25 | 2,395 | 1.26 | 1.15 | 15.10 |
| | AII | 937 | 0.79 | 0.88 | 6.90 | 2,554 | 1.05 | 1.56 | 22.30 |
| IM | JII | 543 | 1.33 | 0.79 | 3.95 | 2,339 | 1.25 | 0.73 | 3.95 |
| | AII | 906 | 0.58 | 0.90 | 6.86 | 3,371 | 0.67 | 1.00 | 21.13 |
| JA | JII | 2,402 | 1.17 | 0.72 | 3.84 | 6,136 | 1.30 | 0.90 | 15.10 |
| | AII | 2,765 | 0.75 | 1.00 | 19.33 | 6,826 | 0.93 | 1.54 | 48.43 |
| KV | JII | 769 | 1.18 | 0.58 | 2.92 | 753 | 1.29 | 0.64 | 2.92 |
| | AII | 785 | 0.97 | 1.17 | 11.04 | 783 | 1.06 | 1.36 | 20.11 |
| KY | JII | 603 | 0.92 | 0.89 | 3.83 | 566 | 1.23 | 1.17 | 13.55 |
| | AII | 751 | 0.73 | 0.95 | 7.13 | 703 | 0.92 | 1.28 | 11.35 |
| LE | JII | 2,804 | 1.19 | 0.73 | 15.69 | 4,108 | 1.19 | 1.00 | 17.37 |
| | AII | 3,067 | 0.88 | 1.04 | 11.24 | 4,867 | 0.88 | 1.55 | 48.43 |
| MU | JII | 236 | 1.02 | 0.51 | 2.41 | 826 | 1.22 | 0.62 | 2.41 |
| | AII | 982 | 0.31 | 0.64 | 5.26 | 1,951 | 0.53 | 0.93 | 9.37 |
| NE | JII | 1,456 | 1.00 | 0.77 | 10.00 | 2,745 | 1.16 | 1.12 | 17.03 |
| | AII | 1,492 | 0.75 | 1.02 | 12.11 | 2,923 | 0.84 | 1.52 | 44.47 |
| NN | JII | 859 | 1.11 | 0.63 | 9.28 | 3,110 | 1.33 | 1.00 | 17.03 |
| | AII | 880 | 0.79 | 1.09 | 17.43 | 3,156 | 0.94 | 1.78 | 44.20 |
| PI | JII | 721 | 1.00 | 0.46 | 2.78 | 2,202 | 1.13 | 0.61 | 14.47 |
| | AII | 738 | 0.78 | 0.74 | 4.86 | 2,222 | 0.94 | 1.17 | 15.82 |
| QQ | JII | 786 | 1.01 | 0.63 | 2.94 | 2,525 | 1.20 | 1.01 | 15.69 |
| | AII | 872 | 0.67 | 0.73 | 6.14 | 2,748 | 1.04 | 1.92 | 39.47 |
| QU | JII | 880 | 1.12 | 0.69 | 10.91 | 4,477 | 1.18 | 0.90 | 17.56 |
| | AII | 888 | 1.02 | 1.41 | 24.17 | 4,505 | 0.97 | 1.44 | 44.20 |
| RE | JII | 427 | 1.00 | 0.67 | 2.51 | 935 | 1.15 | 0.57 | 3.58 |
| | AII | 441 | 0.70 | 0.74 | 7.18 | 951 | 0.86 | 0.96 | 8.97 |
| TE | JII | 502 | 0.92 | 0.58 | 2.72 | 490 | 1.16 | 0.83 | 13.55 |
| | AII | 525 | 0.73 | 0.68 | 4.15 | 500 | 0.94 | 1.10 | 12.75 |
| ZC | JII | 994 | 0.97 | 0.75 | 4.23 | 4,127 | 0.91 | 0.72 | 4.88 |
| | AII | 1,022 | 0.82 | 1.08 | 8.55 | 4,164 | 0.64 | 1.07 | 23.70 |
| ZM | JII | 661 | 0.68 | 0.44 | 3.42 | 2,141 | 0.91 | 0.72 | 17.37 |
| | AII | 666 | 0.58 | 0.60 | 4.17 | 2,151 | 0.78 | 0.99 | 11.96 |
| ZR | JII | 1,049 | 1.14 | 0.63 | 2.85 | 1,998 | 1.24 | 0.64 | 4.19 |
| | AII | 1,217 | 0.72 | 0.86 | 8.37 | 2,298 | 0.85 | 1.14 | 12.58 |

* DE=Plant Sciences; DM=Oncology; GC=Geochemistry & Geophysics; GU=Ecology; IM=Engineering, Civil; JA=Environmental Sciences; KV=Geography, Physical; KY=Geology; LE=Geosciences, Multidisciplinary; MU=Horticulture; NE=Public, Environmental & Occupational Health; NN=Infectious Diseases; PI=Marine & Freshwater Biology; QQ=Meteorology & Atmospheric Sciences; QU=Microbiology; RE=Mineralogy; TE=Paleontology; ZC=Veterinary Sciences; ZM=Zoology; ZR=Water Resources
** JII = field-normalized IF; AII = field-normalized citation

### 3.2.2 Two-sample Wilcoxon rank-sum (Mann-Whitney) test

Next, we verified if the differences observed in the distributions of each impact factor are significant. Having verified by the Shapiro-Wilk test, that the distribution of the indicators JIR, JII, AIR and AII are not normal, we then carried out the comparison of the two subsets using the non-parametric Wilcoxon-Mann-Whitney test. This test does not make any assumptions about the distribution. Furthermore we draw no samples from the two subsets - avoiding the problem of randomness of the samples - and the two subsets are mutual independent. For the calculations concerning JIR and JII, we again exclude consideration of the proceedings papers. Out of a total of 80 comparisons (20 per each of the four indicators), the test reveals statistically significant differences between the "Country" and "No country" subsets in 71 cases (88.8%).

In the remaining nine cases (11.3%), the differences result as not significant. These instances occur as follows:
- For the indicator JIR, in IM_Engingeering, civil;
- For indicator JII, in IM_Engineering, civil and QU_Microbiology;
- For AIR, in DM_Oncology;
- For AII, in DM_Oncology; KV_Geography, physical; KY_Geology; NN_Infectious diseases; ZR_Water resources.

There are also several subject categories where there are greater impact values for the Country subset than for No country, but the corresponding differences are not observed as significant. These are in IM_Engineering, civil, for the indicators JIR and JII, and in DM_Oncology for the indicator AIR.

### 3.2.3 Rate of concentration of highly cited articles (HCAs)

Finally, we extracted the publications with AIR above $95^{th}$ percentile (HCAs) in the two subsets "Country" and "No country". Then we measured the share of HCAs in each subject category. The data in Table 9 indicate that the share of HCAs in the Country subset is higher than that recorded in the No country subset in only two subject categories: IM_Engineering, civil and ZC_Veterinary sciences.

In the same manner as the preceding analysis, we then extracted the publications in "high impact journals" (PHIJ), identified on the basis of the JIR indicator being higher than $95^{th}$ percentile. In this case the dataset again excludes the proceedings papers. The results presented in Table 10 show that the share of PHIJ in the Country subset is higher than that registered in the No country subset in only one subject category (ZM_Zoology).

*Table 9: Analysis of concentration of highly cited articles (by indicator AIR) for the 20 subject categories (row percentage in brackets)*

| Subject Category* | HCAs "Country" | HCAs "No country" |
|---|---|---|
| DE | 17 out of 1,253 (1.36%) | 262 out of 4,353 (6.02%) |
| DM | 45 out of 1,304 (3.45%) | 863 out of 15,675 (5.51%) |
| GC | 76 out of 1,801 (4.22%) | 250 out of 3,472 (7.2%) |
| GU | 25 out of 937 (2.67%) | 180 out of 2,554 (7.05%) |
| IM | 37 out of 906 (4.08%) | 132 out of 3,371 (3.92%) |
| JA | 95 out of 2,765 (3.44%) | 408 out of 6,826 (5.98%) |
| KV | 43 out of 785 (5.48%) | 56 out of 783 (7.15%) |
| KY | 30 out of 751 (3.99%) | 53 out of 703 (7.54%) |
| LE | 147 out of 3,067 (4.79%) | 254 out of 4,867 (5.22%) |
| MU | 15 out of 982 (1.53%) | 48 out of 1,951 (2.46%) |
| NE | 49 out of 1,492 (3.28%) | 164 out of 2,923 (5.61%) |
| NN | 21 out of 880 (2.39%) | 165 out of 3,156 (5.23%) |
| PI | 18 out of 738 (2.44%) | 116 out of 2,222 (5.22%) |
| QQ | 18 out of 872 (2.06%) | 201 out of 2,748 (7.31%) |
| QU | 45 out of 888 (5.07%) | 237 out of 4,505 (5.26%) |
| RE | 10 out of 441 (2.27%) | 32 out of 951 (3.36%) |
| TE | 8 out of 525 (1.52%) | 26 out of 500 (5.2%) |
| ZC | 59 out of 1,022 (5.77%) | 117 out of 4,164 (2.81%) |
| ZM | 4 out of 666 (0.6%) | 66 out of 2,151 (3.07%) |
| ZR | 42 out of 1,217 (3.45%) | 120 out of 2,298 (5.22%) |

\* DE=Plant Sciences; DM=Oncology; GC=Geochemistry & Geophysics; GU=Ecology; IM=Engineering, Civil; JA=Environmental Sciences; KV=Geography, Physical; KY=Geology; LE=Geosciences, Multidisciplinary; MU=Horticulture; NE=Public, Environmental & Occupational Health; NN=Infectious Diseases; PI=Marine & Freshwater Biology; QQ=Meteorology & Atmospheric Sciences; QU=Microbiology; RE=Mineralogy; TE=Paleontology; ZC=Veterinary Sciences; ZM=Zoology; ZR=Water Resources

*Table 10: Analysis of the concentration of publications in high impact journals (by indicator JIR) for the 20 subject categories (row percentage in brackets)*

| Subject Category* | PHIJ "Country" | PHIJ "No country" |
|---|---|---|
| DE | 6 out of 928 (0.65%) | 341 out of 3,868 (8.82%) |
| DM | 94 out of 1,299 (7.24%) | 1428 out of 15,608 (9.15%) |
| GC | 112 out of 1,688 (6.64%) | 325 out of 3,311 (9.82%) |
| GU | 21 out of 846 (2.48%) | 129 out of 2,395 (5.39%) |
| IM | 16 out of 543 (2.95%) | 118 out of 2,339 (5.04%) |
| JA | 127 out of 2,402 (5.29%) | 633 out of 6,136 (10.32%) |
| KV | 29 out of 769 (3.77%) | 58 out of 753 (7.7%) |
| KY | 47 out of 603 (7.79%) | 102 out of 566 (18.02%) |
| LE | 81 out of 2,804 (2.89%) | 188 out of 4,108 (4.58%) |
| MU | 0 out of 236 (0%) | 0 out of 826 (0%) |
| NE | 67 out of 1,456 (4.6%) | 277 out of 2,745 (10.09%) |
| NN | 12 out of 859 (1.4%) | 134 out of 3,110 (4.31%) |
| PI | 3 out of 721 (0.42%) | 17 out of 2202 (0.77%) |
| QQ | 30 out of 786 (3.82%) | 373 out of 2,525 (14.77%) |
| QU | 15 out of 880 (1.7%) | 137 out of 4,477 (3.06%) |
| RE | 7 out of 427 (1.64%) | 18 out of 935 (1.93%) |
| TE | 9 out of 502 (1.79%) | 27 out of 490 (5.51%) |
| ZC | 19 out of 994 (1.91%) | 84 out of 4,127 (2.04%) |
| ZM | 3 out of 661 (0.45%) | 7 out of 2,141 (0.33%) |
| ZR | 25 out of 1,049 (2.38%) | 108 out of 1,998 (5.41%) |

\* DE=Plant Sciences; DM=Oncology; GC=Geochemistry & Geophysics; GU=Ecology; IM=Engineering, Civil; JA=Environmental Sciences; KV=Geography, Physical; KY=Geology; LE=Geosciences, Multidisciplinary; MU=Horticulture; NE=Public, Environmental & Occupational Health; NN=Infectious Diseases; PI=Marine & Freshwater Biology; QQ=Meteorology & Atmospheric

*Sciences; QU=Microbiology; RE=Mineralogy; TE=Paleontology; ZC=Veterinary Sciences; ZM=Zoology; ZR=Water Resources*

## 4. Discussion and conclusions

The literature generally agrees that there are certain "non-content related factors" that affect the visibility and citability of publications, and these concern all scientists in their attempts to diffuse their results to the larger community (Jamali et al., 2014). It is no surprise then that a number of scientometricians have investigated the non-content factors that could affect the visibility or influence of a scientific article. Among these factors, much attention has been given to the structure of the publication titles. An appropriate title permits not only greater traceability and recovery in bibliographic searches, but also higher citability. The present work has examined whether the presence of a term referring to a country in the title, abstract or keywords could be a penalizing factor in terms of visibility and impact in the relevant scientific community.

The results of the analysis provide general confirmation of the initial hypothesis. In particular, this holds true for articles, reviews, and conference proceedings, while it is not so clear for letters. A similar comparison between the publications with the country's name in the title only and those without shown that the impact differences are even higher than in the previous analysis.

The analyses at the deeper, sectoral level, shown for a selected number of subject categories (chosen based on greater shares of products that include country's name), demonstrate that the values of the impact indicators for the products of the "*No country*" subset are systematically greater than those for the "*Country*" subset, with the exception of a limited number of subject categories, although not for all the indicators considered. The final in-depth examination of highly-cited articles reveals that the share of HCAs in the No country subset is greater than that registered in the Country subset, in all but two of the 20 subject categories considered. Similarly, the share of articles published in high impact journals in the No country subset is greater than that registered in the Country subset in all but one subject category.

The reasons underlying this phenomenon are easily intuited. To the international community of researchers confronting the literature, the studies conducted at the country level would typically be less appealing that those dealing with the same subjects at the broader level. The researcher could suspect that certain results would be influenced by country-specific traits, and therefore be difficult to generalize.

Future research may examine whether differences in impact occur among countries cited in the title. Whatever the reasons for the phenomenon, we avoid the practice of some of our previous publications: specifically, we now make no reference to Italy in the titles or keywords of our work!